\documentclass[apj,iop]{emulateapj}

\pdfoutput=1

\usepackage{amsmath}
\usepackage{amssymb}
\usepackage{latexsym}
\usepackage{wasysym}
\usepackage{url}
\usepackage{ulem}

\shorttitle{The Logarithmic Density-Displacement Relation}
\shortauthors{Falck et al.}

\begin{document}

\newcommand{\grad}{\nabla}
\newcommand{\figscl}{0.34}
\newcommand{\bPsi}{\mathbf{\Psi}}

\title{Straightening the Density-Displacement Relation with a Logarithmic Transform}
\author{Bridget L. Falck\altaffilmark{1}, Mark C. Neyrinck\altaffilmark{1}, Miguel A. Aragon-Calvo\altaffilmark{1}, Guilhem Lavaux\altaffilmark{1,2}, and Alexander S. Szalay\altaffilmark{1}}
\altaffiltext{1}{Department of Physics and Astronomy, Johns Hopkins University, 3400 N Charles St, Baltimore, MD 21218, USA}
\altaffiltext{2}{Department of Physics, University of Illinois at Urbana-Champaign, 1110 West Green St, Urbana, IL 61801, USA}

\begin{abstract}
We investigate the use of a logarithmic density variable in estimating the Lagrangian displacement field, motivated by the success of a logarithmic transformation in restoring information to the matter power spectrum. The logarithmic relation is an extension of the linear relation, motivated by the continuity equation, in which the density field is assumed to be proportional to the divergence of the displacement field; we compare the linear and logarithmic relations by measuring both of these fields directly in a cosmological $N$-body simulation. 
The relative success of the logarithmic and linear relations depends on the scale at which the density field is smoothed. Thus we explore several ways of measuring the density field, including Cloud-In-Cell smoothing, adaptive smoothing, and the (scale-independent) Delaunay tessellation, and we use both a Fourier space and a geometrical tessellation approach to measuring the divergence. 
We find that the relation between the divergence of the displacement field and the density is significantly tighter with a logarithmic density variable, especially at low redshifts and for very small ($\sim$2 h$^{-1}$Mpc) smoothing scales. We find that the grid-based methods are more reliable than the tessellation-based method of calculating both the density and the divergence fields, though in both cases the logarithmic relation works better in the appropriate regime, which corresponds to nonlinear scales for the grid-based methods and low densities for the tessellation-based method.
\end{abstract}


\section{Introduction}

Under the influence of gravity, the initially-small fluctuations in the primordial density field grow to become the hierarchical and nonlinear structures observed today. We can study the late-time clustering of matter by evolving a density field described by an initial power spectrum and comparing the simulated structures to observed galaxies in a statistical way. However, if we were able to undo the smearing effects of nonlinear evolution and reverse the formation of structure, we would be able to reconstruct the primordial power spectrum from an observed distribution of galaxies.
Many reconstruction methods have been developed~\citep{ham91,wei92,pea94,pea96,fri02,bre03,moh06}, from simple linear reconstruction to more complex methods, with varying degrees of success (see~\citep{nar99} for a comparative review).
For example, Monge-Amp\`{e}re-Kantorovitch (MAK) reconstruction \citep{fri02,bre03,moh06} was successfully able to reproduce the observed velocity field of the Local Group~\citep{lav10}.
A reconstruction method that uses the linear Zel'dovich approximation of the Lagrangian displacement has also been proposed to enhance the measurement of baryon acoustic oscillations~\citep[BAOs;][]{eisrec,noh09,pad09,whi10,meh11}. The displacement field is a crucial quantity in most of these methods, and is of interest in its own right.

There is reason to believe that a logarithmic transformation of the density field may aid estimation of the displacement field.
The mass distribution has been successfully described by a lognormal field that evolved from Gaussian initial conditions \citep{coles91}, with some evidence for a skewed lognormal field in the nonlinear regime \citep{col94}. More recently, it has been found that the power spectrum of the log-transformed density field contains more Fisher information than the usual power spectrum at small scales by up to a factor of $\sim$10 \citep{ney09}. A modified logarithmic transform has also been shown to increase the precision of the power spectrum of the nonlinear weak lensing convergence field \citep{seo10}. The log-transformed density field is also more effective in constraining cosmological parameters than the standard density field when using the power spectrum~\citep{ney11}.

This paper investigates the effect of a logarithmic transform of the density field on the relation between the density and displacement fields. We do this by measuring the divergence of the displacement field, which in linear theory is proportional to the negative density contrast $\delta$, using both Eulerian and Lagrangian techniques in a cosmological $N$-body simulation. We compare the linear and logarithmic approaches and evaluate their dependence on redshift and smoothing scale, both of which affect the applicability of linear theory. The linear and logarithmic approximations for the displacement field are derived in Section~\ref{sec:reconstruction}, and three methods of measuring the divergence of the displacement field and the density field are presented in Section~\ref{sec:results}. Concluding remarks are given in Section~\ref{sec:conclusion}.

\section{Theory}
\label{sec:reconstruction}

In this section, we compare the linear approximation of the continuity equation to a logarithmic approximation. The large-scale dynamical evolution of structure is described by the Zel'dovich approximation \citep{zel70},
\begin{equation}
\mathbf{x} = \mathbf{q} + \mathbf{\Psi} = \mathbf{q} - D\,\nabla\mathbf{\Phi},
\end{equation}
where ${\bf x}$ is the comoving Eulerian coordinate, ${\bf q}$ the Lagrangian coordinate, $\bPsi$ the displacement, $D$ the linear growth function, and $\mathbf{\Phi}$ the gravitational potential. The linear Poisson equation relates the density contrast, $\delta = (\rho - \bar{\rho})/\bar{\rho}$, to the gravitational potential by $\delta = D\,\grad^2\mathbf{\Phi}$, thus the divergence of the linear Zel'dovich displacement can be related to the density contrast:
\begin{equation}
\label{eqn:divdel}
\grad\cdot\,\bPsi = -\delta. 
\end{equation}
The time derivative of the Zel'dovich approximation in expanding coordinates (${\bf r} = a{\bf x}$, where ${\bf r}$ is the physical coordinate) allows us to relate this expression to the continuity equation. We first note that the derivative consists of two components, $\dot{\bf r} = \dot{a}{\bf x} + a\dot{\bf x}$, where the first component describes the rate of expansion and the second the peculiar velocity of galaxies (or mass particles): ${\bf v} = a\dot{\bf x} = a\dot\bPsi$. The continuity equation can be written as~\citep[see, e.g.,][p 48]{pee80}
\begin{equation}
\label{eqn:continuity}
\frac{\partial\left(1+\delta\right)}{\partial t} + \frac{1}{a}\grad\cdot[\left(1 + \delta\right){\bf v}] = 0.
\end{equation}
To first in $\delta$ order this becomes
\begin{equation}
\frac{\partial\delta}{\partial t} + \frac{1}{a}\grad\cdot\,{\bf v} = 0,
\end{equation}
thus we have $\dot\delta = -\grad\cdot\,\dot\bPsi$ and have recovered Equation~\ref{eqn:divdel}.

The continuity equation (Eqn.~\ref{eqn:continuity}) may also be simplified by keeping the $(1+\delta)$ together, resulting in a logarithmic derivative of $(1+\delta)$:
\begin{equation}
\frac{1}{\left(1+\delta\right)}\frac{\partial\left(1+\delta\right)}{\partial t} + \frac{1}{a}\grad\cdot\,{\bf v} = 0.
\end{equation}
This gives a logarithmic expression for the divergence of the Lagrangian displacement,
\begin{equation}
\label{eqn:divlnd}
\grad\cdot\,\bPsi = - \ln\left(1 + \delta\right)+C,
\end{equation}
where the $C$ term appears because $\ln(1+\delta)$ is not a zero-mean-field, so we must take into account a non-zero integration constant. We find that this constant is well-approximated by the rather natural value of $\langle\ln(1+\delta)\rangle$. We now may compare the linear relation, given by Eqn.~\ref{eqn:divdel}, to the logarithmic relation, given by Eqn.~\ref{eqn:divlnd}.

\section{Results}
\label{sec:results}

We measure the density field and $\grad\cdot\bPsi$ in a 200~h$^{-1}$Mpc, $256^3$ particle Gadget~\citep{spr01} cold dark matter simulation with standard $\Lambda$CDM cosmology ($\Omega_M = 0.3, \Omega_\Lambda = 0.7, h = 0.7$). We use three different methods: a grid-based Cloud-In-Cell (CIC) smoothing plus a Fourier-space estimation of the divergence (Section~\ref{sec:cic}); an adaptive smoothing using a Smoothed Particle Hydrodynamics (SPH) kernel, combined with a Fourier-space divergence calculation (Section~\ref{sec:sph}); and a geometrical Delaunay tessellation estimation of both the density and the divergence (Section~\ref{sec:dtfe}).

We find that each of these methods achieves differing degrees of success based on the nature of the length-scales involved: the CIC method is an Eulerian, mass-weighted scheme; the Delaunay tessellation is a Lagrangian, volume-weighted scheme; and the adaptive mesh method is a hybrid, featuring a Lagrangian kernel with interpolation onto an Eulerian grid. We discuss each in turn below.

\subsection{Cloud-In-Cell and FFT}
\label{sec:cic}

The CIC method smooths an arbitrary distribution of particles onto a regularly-spaced grid by placing a ``cloud'' having the volume of a grid cell around each particle, so that a particle contributes to the average of multiple nearby cells based on the fraction of its volume contained in these cells. Using this weighting scheme, we calculate the density and displacement fields in real space for both $64^3$ and $128^3$ cells, where the displacement $\bPsi$ is the final ($z=0$) minus initial ($z=49$) particle positions. In the case of $128^3$ cells at $z=0$, there is only one cell containing zero particles; we set its value to the average value in the surrounding cells. To avoid cells with zero particles, this is the finest grid we use, given the resolution of the simulation.
The divergence of the displacement can then easily be calculated in Fourier space, where the derivatives become simple multiples of $k$.

We note here that BAO reconstruction methods also calculate the linear Zel'dovich displacement in Fourier space, which they multiply by a smoothing function $S(k)$: $\bPsi_{k,\rm{lin}} = i\,{\bf k}\,\delta_k\,S(k)/k^2$; for example, a Gaussian smoothing function would take the form $S(k) = e^{-k^2R^2/2}$. We do not apply an extra smoothing function in Fourier space, but note that the size of the grid cells gives the effective smoothing length of the both the density and displacement fields. In our simulations, a $64^3$ cell grid has a length of 3.1~h$^{-1}$Mpc and a $128^3$ cell grid has length 1.6~h$^{-1}$Mpc, while the reconstruction methods have $R$ = 5, 10, and 20~h$^{-1}$Mpc, with $R$=10~h$^{-1}$Mpc performing the best \citep{eisrec,noh09}.

\begin{figure}[tb]
\centering
\includegraphics[scale=\figscl]{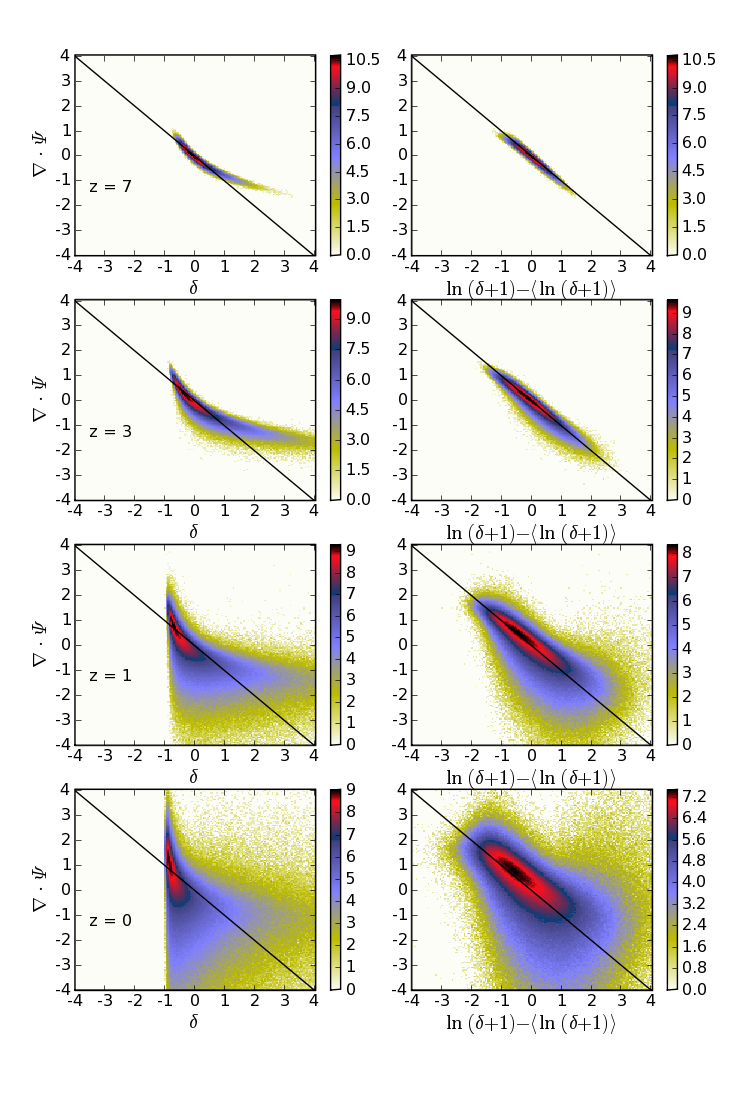}
\caption{The divergence of the Lagrangian displacement as a function of $\delta$ (left) and $\ln\left(1+\delta\right)$ (right), calculated using CIC smoothing for a cell size of 1.6 h$^{-1}$Mpc at 4 different redshifts. A slope of -1 through the origin is plotted for reference. The color scale is logarithmic and gives the number of CIC cells in each histogram bin.\label{fig:divcic4z}}
\end{figure}

In Figure~\ref{fig:divcic4z} we show a two-dimensional histogram of $\grad\cdot\bPsi$ versus $\delta$ and $\ln(1+\delta)$ at four different redshifts, for a cell size of 1.6 h$^{-1}$Mpc and with a slope of -1 plotted for reference. The color scale is logarithmic so that the outlying cells can be seen, though the furthest outliers (especially at low-$z$) extend beyond the range of the plot. At $z=7$, the nonlinear clustering of matter has only begun for the initially highest-density peaks, so both the linear and logarithmic relations have slopes near -1 and are very tight. At $z=3$ there is more scatter, and the linear relation begins to deviate very slightly from a slope of -1 for the majority of cells (shown in black in the plot) and to develop a high $\delta$ tail. This trend continues at $z=1$, with increased scatter in both the logarithmic and linear relations, and with the linear relation deviating from a slope of -1. Finally, by $z=0$ the linear relation deviates drastically from a slope of -1 while the logarithmic relation does very well, though with much scatter.

\begin{figure}[tb]
\centering
\includegraphics[scale=\figscl]{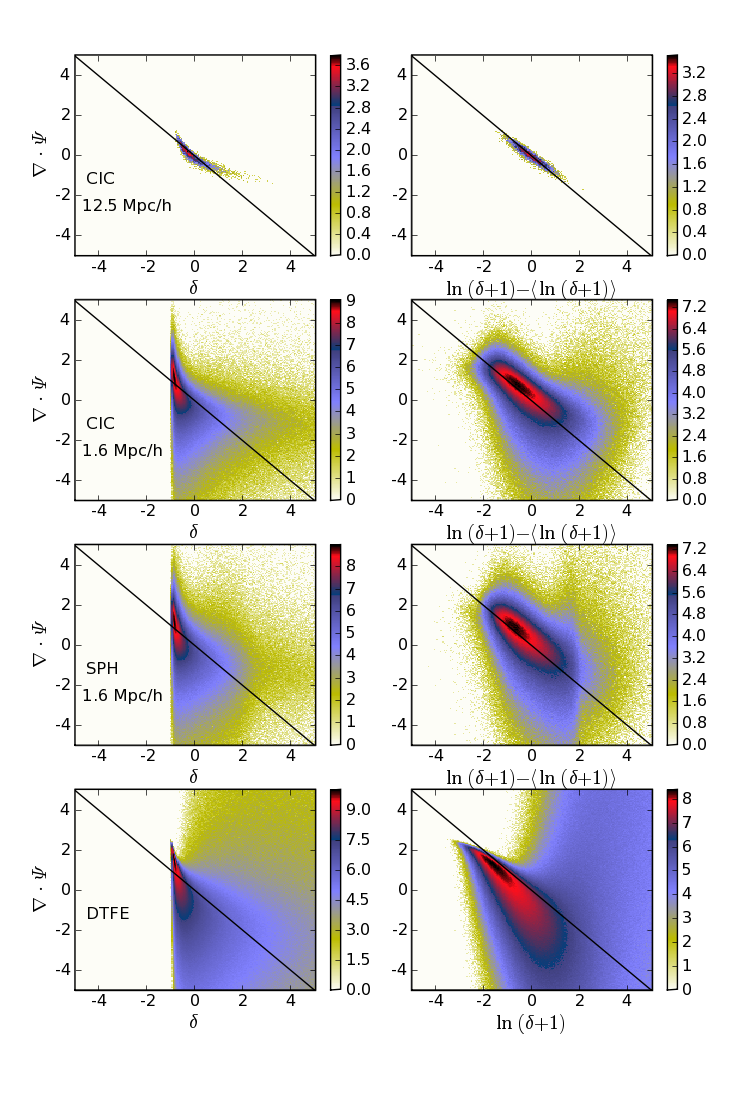}
\caption{The divergence of the Lagrangian displacement as a function of $\delta$ (left) and $\ln\left(1+\delta\right)$ (right) at $z=0$, calculated using CIC smoothing for a cell size of 12.5 h$^{-1}$Mpc (top) and 1.6 h$^{-1}$Mpc (upper middle), adaptive SPH-like smoothing with a cell size of 1.6~h$^{-1}$Mpc (lower middle), and the Delaunay Tessellation Field Estimator (DTFE) method (bottom). A slope of -1 through the origin is plotted for reference. The color scale is logarithmic and gives the number of cells (CIC and SPH) or particles (DTFE) in each histogram bin.\label{fig:divz0}}
\end{figure}

It is not unexpected that the linear relation does not hold, since 1.6~h$^{-1}$Mpc is by no means in the linear regime, but the fact that the logarithmic relation holds so well echoes the results of \citet{ney09} and others that the logarithmic transform of the density field increases the information content of the matter power spectrum in the weakly nonlinear regime. 
When the CIC cell size is increased to 12.5~h$^{-1}$Mpc, both the linear and logarithmic relations become very tight, though the logarithmic approximation continues to do slightly better.
Figure~\ref{fig:divz0} is the same as Figure~\ref{fig:divcic4z} but at $z=0$, showing the CIC method for cell sizes of 1.6 h$^{-1}$Mpc and 12.5~h$^{-1}$Mpc, along with the two other methods discussed in the next sections.

\subsection{Adaptive SPH Smoothing and FFT}
\label{sec:sph}

We also try an adaptive smoothing technique, described by \citet{col07}, on both the density and the displacement fields. We assume that each particle $p$ traces some physical quantity $A_p$. For the purpose of this paper, $A_p$ will be either a particle count or the displacement vector $\mathbf{\Psi}$ between the final and initial particle positions.
We sample the adaptively-smoothed field onto a three-dimensional grid, with grid sites denoted by the integers $(i,j,k)$. At each grid site we define the smoothing radius $R(i,j,k)$ which corresponds to half the distance from the grid site to the $M$-th particle, where $M$ is defined as:
\begin{equation}
	M(i,j,k) = \max\left\{M_\text{min}, M_\mathrm{binned}(i,j,k)\right\}
\end{equation}
and $M_\mathrm{binned}(i,j,k)$ is the number of particles binned on the grid site $(i,j,k)$. Following~\citet{col07}, we conservatively set $M_\text{min}=32$, corresponding to the typical number of a particle's closest neighbors. This choice of $M(i,j,k)$ ensures that we always take into account a sufficient number of particles for making the adaptively-smoothed field on the grid.

The smoothed-interpolated value $\tilde{A}(i,j,k)$ at the grid site $(i,j,k)$, which corresponds to a smooth-averaging of the quantities $A_p$ carried out by individual particles $p$, is defined as:
\begin{equation}
	\tilde{A}_{i,j,k} = \frac{1}{R(i,j,k)^3} \sum_{p=1}^{N} A_p W_p \mathcal{S}\left(\frac{d_p}{R(i,j,k)}\right),
\end{equation}
where $W_p$ are additional weights to be determined for each particle and $\mathcal{S}$ is the smoothing kernel function. We determine the weights by enforcing that the sum of $\tilde{A}_{i,j,k}$ over all grid sites must be the same as the sum of $A_p$ over all particles. This yields, for $N$ particles,
\begin{multline}
	\sum_{i,j,k} \tilde{A}_{i,j,k} = \sum_{i,j,k} \frac{1}{R(i,j,k)^3} \sum_{p=1}^{N} A_p W_p \mathcal{S}\left(\frac{d_p}{R(i,j,k)}\right) = \\ \sum_{p=1}^N A_p = \sum_{p=1}^N A_p W_p \sum_{i,j,k} \frac{1}{R(i,j,k)^3} \mathcal{S}\left(\frac{d_p}{R(i,j,k)}\right).
\end{multline}
The last equality is true if
\begin{equation}
	W_p = \left(\sum_{i,j,k} \frac{1}{R(i,j,k)^3} \mathcal{S}\left(\frac{d_p}{R(i,j,k)}\right)\right)^{-1}.
\end{equation}
Finally, we define the smoothing kernel to be the typical SPH compact-support spline~\citep{mon92},
\begin{equation}
	\mathcal{S}(x) = \left\{ 
			\begin{array}{ll}
				1 - \frac{3}{2} x^2 + \frac{3}{4} x^3 & 0 \leq x \leq 1 \\
				\frac{1}{4}(2-x)^3 & 1 \leq x \leq 2 \\
				0 & x \geq 2
			\end{array}.\right.
\end{equation}

We now compute both the mass density fluctuation field $\delta(i,j,k)$ and the displacement field $\mathbf{\Psi}(i,j,k)$ in Eulerian coordinates. In the case of $N$-body simulations, the mass density fluctuation field is obtained by putting a mass $m_p=1$ on each particle $p$, thus
\begin{equation}
	\delta(i,j,k) = \frac{\tilde{m}(i,j,k)}{N/N_\text{g}} - 1,
\end{equation}
where we used the conservation of the quantity $\sum m_p=\sum \tilde{m}(i,j,k)$, and defined $N_\text{g}$ the number of grid sites. In the case of observations, one could in principle use any tracer of the mass that is available, such as the luminosity or the number of galaxies. 
The displacement field is obtained by averaging the displacement carried by each of the tracer particles. We thus obtain
\begin{equation}
	\mathbf{\Psi}(i,j,k) = \frac{\tilde{\Psi}(i,j,k)}{\tilde{1}(i,j,k)}
\end{equation}
where $\tilde{1}$ is the result of applying the adaptive filter to compute the number of tracers on the grid site $(i,j,k)$. We note that even though in $N$-body simulations $\tilde{1}(i,j,k)=\tilde{m}(i,j,k)$, this is not the case for observations.

We obtain the divergence of the displacement field in Fourier space, as with the previous CIC method.
Figure~\ref{fig:divz0} shows $\grad\cdot\bPsi$ as a function of $\delta$ and $\ln(1+\delta)$ using the adaptive SPH smoothing method with a cell size of 1.6 h$^{-1}$Mpc (lower middle plot).
We note that both the CIC and SPH methods produce very similar results. The additional advantage of the SPH method is that grid cells with zero particles are not an issue, as they are for CIC, since $M_\text{min}=32$ particles are always included in the binning. 
The vertical artifact near $\ln(1+\delta)-\langle\ln(1+\delta)\rangle=2$ marks the transition where $M_\text{min}$ becomes important, above which the adaptive method behaves effectively the same as the CIC method. Note however that this is a very small feature that is exaggerated by the logarithmic color scaling on the plot.

\subsection{Delaunay Tessellation}
\label{sec:dtfe}

The Voronoi tessellation and its dual, the Delaunay tessellation, divide an $N$-dimensional space into a fully volume-covering set of cells based on the underlying particle distribution (see e.g. \citet{van94}). The resulting distribution of cells is such that the volume of the cells is inversely proportional to the density of the particle field at the location of each sampled particle, leading to a natural and adaptive density estimator~\citep[the Delaunay Tessellation Field Estimator, DTFE;][]{sch00,pel03,van09}. In the case of the Delaunay tessellation, each particle is surrounded by a set of tetrahedra of which it makes up one of the vertices. The volume of one tetrahedron, $T$, is a function of the positions of the three other vertices for a particle at the origin:
\begin{equation}
V_T = \frac{1}{8}\,{\bf x_1}\cdot\left({\bf x_2}\times{\bf x_3}\right).
\end{equation}
The density is then given by the inverse of the volume of the set of tetrahedra surrounding each particle, $V = \sum_T V_T$. 

\begin{figure}[htb]
\centering
\includegraphics[angle=90,scale=.4]{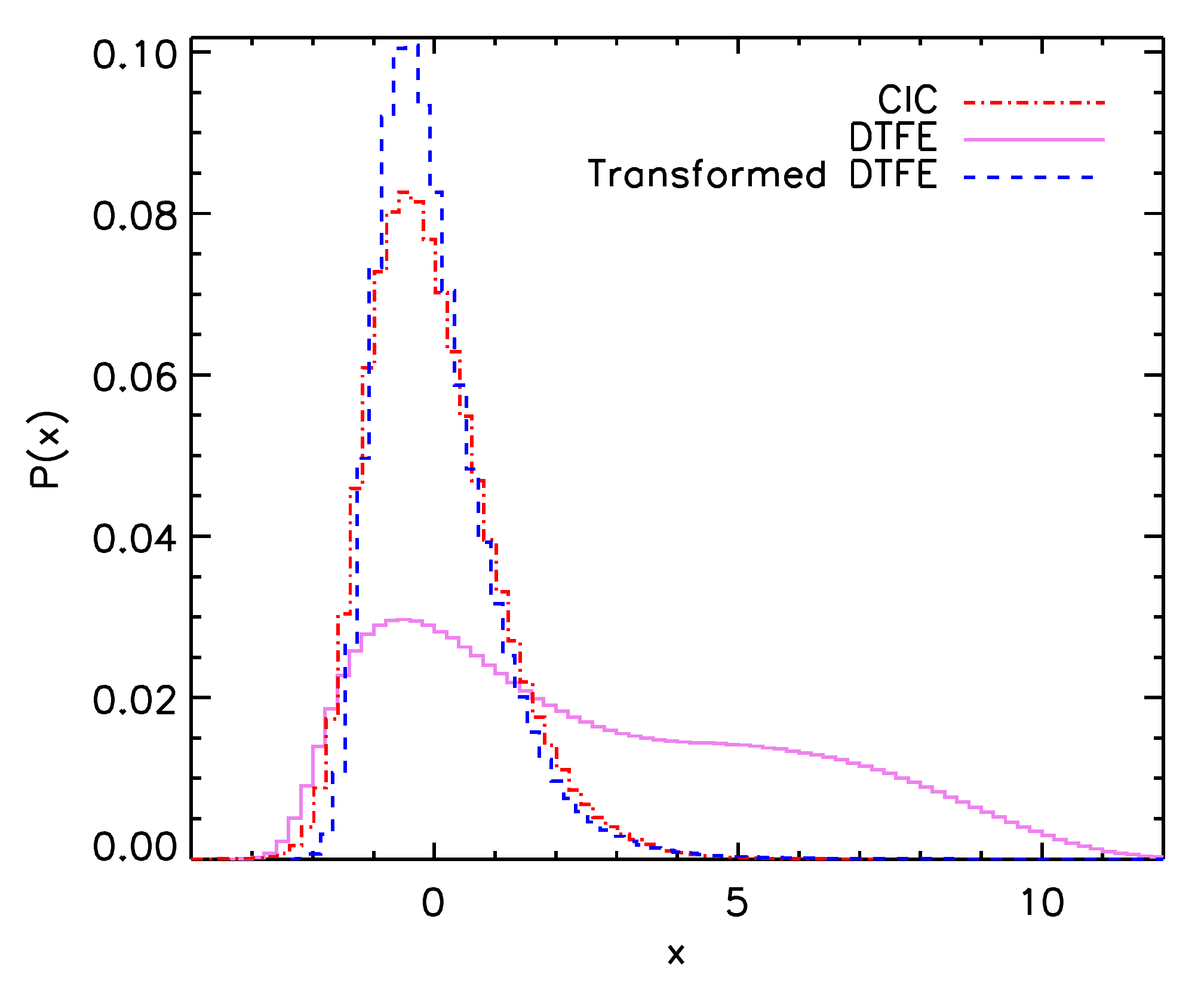}
\caption{Distribution functions of $\ln(1+\delta)-\langle\ln(1+\delta)\rangle$ for the CIC method with 1.6 h$^{-1}$Mpc cell sizes, $\ln(\langle V \rangle / V)$ for the DTFE-calculated volume $V$, and the volume-weighted DTFE transformation (see the text for explanation). The CIC and transformed DTFE distributions are shifted to have zero mean, while the DTFE distribution is not. The good match between the CIC and transformed DTFE distribution functions motivate our use of $\delta = \langle V \rangle / V - 1$ as the DTFE density variable. \label{fig:pdf}}
\end{figure}

We choose to normalize the inverse volumes by $1/\langle V \rangle$ instead of $\langle 1/V \rangle$ (which gives a mean-zero density field) because the DTFE density distribution has a prominent tail at high densities. This can be seen in Figure~\ref{fig:pdf}, in which we plot the distribution functions of the CIC $\ln(1+\delta)-\langle\ln(1+\delta)\rangle$ and the DTFE $\ln(\langle V \rangle / V) = \ln(1+\delta)$. The CIC distribution is much more symmetrical than the DTFE, which has a shoulder at high densities. 
Note that the mean of the DTFE distribution has not been subtracted so that its peak is much closer to zero. We also plot a transformed DTFE distribution function in which we weight the original distribution function by the volume: $P(x) = (V/\langle V \rangle)P(\ln(\langle V \rangle / V))$; this effectively transforms the (mass-weighted) Lagrangian DTFE distribution into a (volume-weighted) Eulerian distribution that closely resembles the CIC distribution.

The Delaunay tessellation also allows a calculation of the divergence of $\bPsi$ at the location of every particle by using Gauss's theorem, $\int_V\grad\cdot\,\bPsi\,dV = \int_S\bPsi\cdot\,{\bf n}\,dS$. Replacing the integrals with discrete sums, we have
\begin{equation}
\grad\cdot\,\bPsi \sum_T V_T = \sum_T \bPsi_T\cdot\,{\bf n}\,dS_T,
\end{equation}
where we take $\bPsi_T$ to be the average of the $\bPsi$ values of the three particles making the outside face of the tetrahedron, and where the surface element is given by
\begin{equation}
dS_T = \frac{1}{2}\,\left({\bf x_1} - {\bf x_2}\right)\times\left({\bf x_1} - {\bf x_3}\right).
\end{equation}
Thus for each particle we have a measure of both $\grad\cdot\bPsi$ and $\delta$.

Figure~\ref{fig:divz0} shows $\grad\cdot\bPsi$ as a function of $\delta$ and $\ln(1+\delta)$ at $z=0$, calculated using the Delaunay tessellation (bottom plot) and the other methods. Both the divergence of $\bPsi$ and the density calculated with the Delaunay tessellation lead to a large scatter at high densities beyond the range of the plot, with the maximum values of $\grad\cdot\bPsi$ reaching around $\pm 1000$. This large scatter is not surprising because neighbors in high-density regions are unlikely to be true Lagrangian neighbors, so their displacements should be fairly random; additionally, the length-scales associated with the highest densities are smaller than the scale over which we would expect the logarithmic approximation to hold. While the low-density success of the logarithmic approximation is encouraging, it seems that a Lagrangian kernel method is not the most appropriate choice for relating the density and displacement fields.

\section{Discussion}
\label{sec:conclusion}

Motivated by the success of the logarithmic density field in improving the statistics of large scale structure into the nonlinear regime, we have compared the linear relation between the density and displacement fields to a logarithmic approach. We have used different methods of measuring the divergence of the displacement field, $\grad\cdot\,\bPsi$, and the density field in cosmological $N$-body simulations, and have investigated their relation for different redshifts and smoothing scales. The logarithmic approach fares better even on scales as small as $\sim$2 h$^{-1}$Mpc, where the linear approximation is no longer valid. Smoothing over a uniform grid in Eulerian coordinates seems to outperform a tessellation-based Lagrangian method, though both work well at low to average densities.

Because a key advantage of the logarithmic transformation is the ability to use the density field on weakly nonlinear scales, this method would be most effective for a low-redshift galaxy survey (such that the effect of nonlinear evolution is strongest) with a high galaxy density (such that the mean separation of galaxies is small and in the nonlinear regime). 
Using a logarithmic transform of the density to reconstruct the displacement field is simpler than advanced algorithms such as MAK reconstruction~\citep{bre03} and thus may require only small changes to existing analysis codes, providing an improvement over the linear method while maintaining a high degree of computational efficiency. Our full investigation of a reconstruction algorithm that uses a logarithmic transform of the density field is left to future work and will be applied to a large suite of cosmological simulations.

When applying this method to an observed galaxy distribution, additional considerations must be accounted for, such as discreteness noise, galaxy bias, and redshift-space distortions. A log-transform can enhance the effective shot noise caused by particle (or galaxy) discreteness, but the log-density can still be usefully estimated in the presence of (low-enough) shot noise.  \citet{ney10} found that even with a rather sparse galaxy sample, added power-spectrum Fisher information in the non-linear regime can be tapped with a log (or similar) transform, suggesting that shot noise can be overcome in the present case, as well.  However, it is likely that higher sampling for the log-density field than
for the overdensity field will be necessary to resolve the fields at a given
resolution.

Though a full reconstruction algorithm using a log-density field remains to be tested, we doubt that a logarithmic transform of the density would exacerbate problems caused by galaxy bias or redshift-space distortions, and these problems seem to be manageable in the case of BAO reconstruction.
For example, it has been shown that linear reconstruction of the BAO peak reduces the effect of bias, including the induced shift in the location of the BAO peak~\citep{meh11}. Additionally,~\citet{eisrec} showed that reconstruction of the BAO peak is effective in both real- and redshift-space.
However, we anticipate that Fingers of God will at some point prevent the push to smaller scales in redshift-space using a logarithmic transform.

We have shown that a simple logarithmic transformation of the density field provides a substantial improvement of the linear theory relation between the density and displacement fields on sub-linear scales. This improvement is greatest at low redshift and for Eulerian smoothing on small ($\sim$2 h$^{-1}$Mpc) scales. This result is in line with the finding that the logarithmic density field restores some lost information to the power spectrum of density fluctuations~\citep{ney09}, and provides more evidence that a logarithmic transform of the density field is a powerful density variable on weakly nonlinear scales.

\acknowledgments

The authors would like to thank the anonymous referee for insightful comments.
BLF and MCN are grateful for support from the Gordon and Betty Moore Foundation. GL acknowledges financial support from NSF Grant AST 07-08849.


\begin{thebibliography}{}
\bibitem[Brenier et al.(2003)]{bre03} Brenier, Y., et al. 2003, MNRAS, 346, 501
\bibitem[Coles \& Jones(1991)]{coles91} Coles, P., \& Jones, B. 1991, MNRAS, 248, 1
\bibitem[Colombi(1994)]{col94} Colombi, S. 1994, ApJ, 435, 536
\bibitem[Colombi et al.(2007)]{col07} Colombi, S., Chodorowski, M. J., \& Teyssier, R. 2007, MNRAS, 375, 348
\bibitem[Eisenstein et al.(2007)]{eisrec} Eisenstein, D. J., Seo, H.-J., Sirko, E., \& Spergel, D. N. 2007, ApJ, 664, 675
\bibitem[Frisch et al.(2002)]{fri02} Frisch, U., Matarrese, S., Mohayaee, R., \& Sobolevski, A. 2002, Nature, 417, 260
\bibitem[Hamilton et al.(1991)]{ham91} Hamilton, A. J. S., Kumar, P., Lu, E., \& Matthews, A. 1991, ApJ, 374, L1
\bibitem[Lavaux et al.(2010)]{lav10} Lavaux, G., Tully, R. B., Mohayaee, R., \& Colombi, S. 2010, ApJ, 709, 483
\bibitem[Mehta et al.(2011)]{meh11} Mehta, K. T., et al. 2011, ApJ, 734, 94
\bibitem[Mohayaee et al.(2006)]{moh06} Mohayaee, R., Mathis, H., Colombi, S., \& Silk, J. 2006, MNRAS, 365, 939
\bibitem[Monaghan(1992)]{mon92} Monaghan, J. J. 1992, ARA\&A, 30, 543
\bibitem[Narayanan \& Croft(1999)]{nar99} Narayanan, V., K., \& Croft, R. A. C. 1999, ApJ, 515, 471
\bibitem[Neyrinck(2011)]{ney11} Neyrinck, M. C. 2011, ApJ, 742, 91
\bibitem[Neyrinck et al.(2009)]{ney09} Neyrinck, M. C., Szapudi, I., \& Szalay, A. S. 2009, ApJ, 698, L90
\bibitem[Neyrinck et al.(2011)]{ney10} Neyrinck, M. C., Szapudi, I., \& Szalay, A. S. 2011, ApJ, 731, 116
\bibitem[Noh et al.(2009)]{noh09} Noh, Y., White, M., \& Padmanabhan, N. 2009, Phys. Rev. D, 80, 123501
\bibitem[Padmanabhan et al.(2009)]{pad09} Padmanabhan, N., White, M., \& Cohn, J. D. 2009, Phys. Rev. D, 79, 063523
\bibitem[Peacock \& Dodds(1994)]{pea94} Peacock, J. A., \& Dodds, S. J. 1994, MNRAS, 267, 1020
\bibitem[Peacock \& Dodds(1996)]{pea96} Peacock, J. A., \& Dodds, S. J. 1996, MNRAS, 280, L19
\bibitem[Peebles(1980)]{pee80} Peebles, P. J. E. 1980, The Large Scale Structure of the Universe (Princeton, NJ: Princeton Univ. Press)
\bibitem[Pelupessy et al.(2003)]{pel03} Pelupessy, F. I., Schaap, W. E., \& van de Wegaert, R. 2003, A\&A, 403, 389
\bibitem[Schaap \& van de Weygaert(2000)]{sch00} Schaap, W. E., \& van de Weygaert, R. 2000, A\&A, 363, L29
\bibitem[Seo et al.(2010)]{seo10} Seo, H.-J., Sato, M., Dodelson, S., Jain, B., \& Takada, M. 2010, ApJ, 729, L11
\bibitem[Springel et al.(2001)]{spr01} Springel, V., Yoshida, N., \& White, S. D. M. 2001, New Astron., 6, 79
\bibitem[van de Weygaert(1994)]{van94} van de Weygaert, R. 1994, A\&A, 283, 361
\bibitem[van de Weygaert \& Schaap(2009)]{van09} van de Weygaert, R., \& Schaap, W. 2009, Data Analysis in Cosmology, 665, 291
\bibitem[Weinberg(1992)]{wei92} Weinberg, D. H. 1992, MNRAS, 254, 315
\bibitem[White(2010)]{whi10} White, M. 2010, arXiv:1004.0250
\bibitem[Zel'dovich(1970)]{zel70} Zel'dovich, Ya. B. 1970, A\&A, 5, 84
\end{thebibliography}
\end{document}